\documentclass[12pt]{article}
\usepackage{amsmath,amssymb,amsthm,color}
\usepackage{graphicx}
\usepackage{cite}
\usepackage{epsfig}
\renewcommand{\baselinestretch}{1.66}

\begin{document}
%
\title{Geometric, magnetic  and electronic properties of folded graphene nanoribbons \\}
\author{
\small Shen-Lin Chang$^{a,b}$, Bi-Ru Wu$^{c,*}$, Po-Hua Yang$^{d}$, Ming-Fa Lin$^{a,*}$ $$\\
\small  $^a$Department of Physics, National Cheng Kung University, Tainan, Taiwan\\
\small  $^b$Department of Physics, National Taiwan Normal University, Taipei, Taiwan\\
\small  $^c$Department of Natural Science, Center for General Education,\\
\small  Chang Gung University, Taoyuan, Taiwan\\
\small  $^d$Department of Mechanical Engineering,\\
\small  Southern Taiwan University of Science and Technology, Tainan 710, Taiwan.\\}
\renewcommand{\baselinestretch}{1.66}
\maketitle

\renewcommand{\baselinestretch}{1.66}
\begin{abstract}
Geometric and electronic properties of folded graphene nanoribbons (FGNRs) are investigated by first-principles calculations. These properties are mainly dominated by the competition or cooperation among stacking, curvature and edge effects. For the zigzag FGNRs, the more stable structures are revealed to be $AB$ stackings, while for the armchair types, $AA^{\prime\prime}$ stackings are more stable. The interlayer interactions and hybridization of four orbitals lead to smaller energy gaps, anti-crossing bands, and more band-edge states. Specifically, the broken mirror symmetry in the odd-$AB$ stacked zigzag FGNRs is responsible for the spin-up and spin-down splitting subbands. All FGNRs are direct-gap semiconductors except that the edge-edge interactions cause the even-$AA$ stacked zigzag FGNRs to exhibit a pair of metallic linear bands. The width-dependent energy gaps in the armchair FGNRs can be classified into six groups. Furthermore, there exist rich features in density of states, including the form, number, intensity and energy of the special structures.

\par\noindent * Corresponding author. {~ Tel:~ +886-6-275-7575;~ Fax:~+886-6-74-7995.}\\~{{\it E-mail address}: brwu@mail.cgu.edu.tw (B.R. Wu); mflin@mail.ncku.edu.tw (M.F. Lin)}

\vskip 1.0 truecm
\par\noindent

\end{abstract}

\pagebreak
\renewcommand{\baselinestretch}{2}
\newpage

{\bf 1. Introduction}
\vskip 0.3 truecm
In 2004, graphene was first successfully synthesized\cite{1}, and since then has attracted considerable attention in the fields of chemistry, materials science and physics\cite{2,3,4,5,6,7,8}. The unique two-dimensional (2D) hexagonal symmetry induces rich electronic properties, such as an extremely high mobility\cite{2}, an anomalous quantum Hall effect\cite{2,6}, and ambipolar transport phenomena\cite{1}. Graphene is viewed as a future electronic material. However, its zero-gap property makes actual applications in logic electronics difficult. In order to further expand the range of applications, opening an energy gap is critical. Semiconducting 1D nanoribbons (GNRs) have been the focus of many experimental\cite{9,10,11,12,13,14,15,16,17,18,19,20,21,22} and theoretical studies\cite{23,24,25,26,27,28,29,30,31,32,33,34,35}. GNRs have been successfully produced from graphene through lithographic techniques\cite{9,10} and from carbon nanotubes by several chemical and physical unzipping methods\cite{11,12,13}. Under different growing environments, GNRs with various structures are presented in laboratories, including few-layer\cite{14}, curved\cite{12}, scrolled\cite{15}, and folded types\cite{16,17,18,19,20}. The unique stacking configuration and curved structure in folded graphene nanoribbons (FGNRs) are worth detailed studies.

Flat GNRs possess rich geometric and electronic properties, being dominated by their edge structures and ribbon widths. Zigzag and armchair GNRs exist according to the achiral boundary condition. A zigzag GNR is a middle-gap semiconductor with two edge-state electrons, which have a ferromagnetic arrangement for each edge but an antiferromagnetic one across the nanoribbon\cite{23}. The zigzag systems are considered as potential materials for future applications in spintronic devices, mainly owing to their halfmetallic characteristics under a transverse electric field\cite{27}. On the other hand, all armchair GNRs are direct-gap semiconductors, which can be classified into three types of energy gaps, being mainly determined by the number of dimer lines ($N_{y}$). The largest (smallest) energy gaps are reported to be at $N_{y}=3p+1$ ($N_{y}=3p+2$); $p$ is a positive integer. All energy gaps are inversely proportional to the ribbon width for all groups\cite{23}.

In previous research studies, the fundamental properties of GNRs with various curvatures or stackings have not been thoroughly explored. Few-layer GNRs\cite{28} present various stacking configurations. The atomic interactions between two layers strongly depend on the van der Waals interactions. The edge-edge interactions\cite{28,29} are introduced by the open edges in some specific configurations. Hence, the electronic properties are dominated by both the stacking- and the edge-related interactions. The hybridization of four orbitals (2$s$,2$p_{x}$,2$p_{y}$,2$p_{z}$) in curved ribbons\cite{24,33} and nanoscrolls causes the complex bondings at the inner side of the curved surface. The larger 2$p_{z}$ orbitals on the outer side of the curved surface provide a compatible environment for adsorbed atoms, such as H and Li\cite{30,31}. As a result, curved systems are promising materials for future hydrogen storage or lithium batteries. Due to the cooperation or competition among the stacking, curvature, and edge effects, a FGNR with curved and stacking parts (Fig. 1) is expected to display intricate and versatile electronic properties. In experiments\cite{16,17,18,19,20}, the synthesized FGNRs have displayed distinct physical characteristics. Experimental measurements from scanning tunneling microcopy show that FGNRs can be formed in various stacking configurations\cite{17,18}. On the other hand, some physical properties of FGNRs have been studied in previous theoretical research\cite{32,33}. However, knowledge of the existence of more stable FGNR configurations, their formation energies, the radius of the curved part, band structures, density of states (DOS), and band gaps is still incomplete. Most research has focused on the $AA$ stacking, but few investigations have been carried out on the more stable $AB$ stackings and other potential configurations.

In this paper, the geometric and electronic properties of zigzag and armchair FGNRs are investigated in detail by first principles calculations. Many stacking configurations are considered, not only on the $AA$ and $AB$ stacking configurations but also $AA^{\prime}$ and $AA^{\prime\prime}$ ones. Among the complex structures, we systematically classify four types of zigzag and armchair systems, based on the edge structure, formation energy, ribbon width and stacking configurations. The armchair systems are further divided into six groups of width-dependent energy gaps on the basis of the stacking configurations. Each of these eight types of FGNRs presents distinct electronic properties, such as a pair of metallic linear bands, metal-semiconductor transitions, splitting of spin-up and spin-down states, and the monotonous width dependence of the energy gaps, etc. Moreover, the predicted main features in DOS are validated and compared to experimental scanning tunneling spectroscopy (STS) measurements. These rich fundamental features in FGNRs can be expected to provide potential applications in energy materials, as well as electronic and spintronic devices.

\vskip 0.6 truecm
\par\noindent
{\bf 2. Theory and geometric structure }
\vskip 0.3 truecm

We use the Vienna ab initio simulation package\cite{36,37} in the density-functional theory (DFT) to investigate the essential properties of FGNRs. The DFT-D2 method\cite{38} is taken into account in order to describe the weak van der Waals interactions. The electron-ion interactions are obtained by the projector-augmented wave method, and the exchange and correlation electron-electron interactions are calculated from the generalized gradient approximation\cite{39} with Perdew-Burke-Ernzerhof. The wave functions are expanded by the plane waves with a maximum kinetic energy of 500 eV. For electronic properties and optimized geometric structures, the first Brillouin zone is, respectively, sampled by 300$\times$1$\times$1 and 12$\times$1$\times$1 k-points in the Monkhorst-Pack scheme. All of the atoms are relaxed until the Hellmann-Feynman force is less than 0.03 eV/${\AA}$. A vacuum space of 15 ${\AA}$ is inserted between periodic images to simulate an isolated FGNR.

\begin{figure}[htbp]
\center
\rotatebox{0} {\includegraphics[width=14cm]{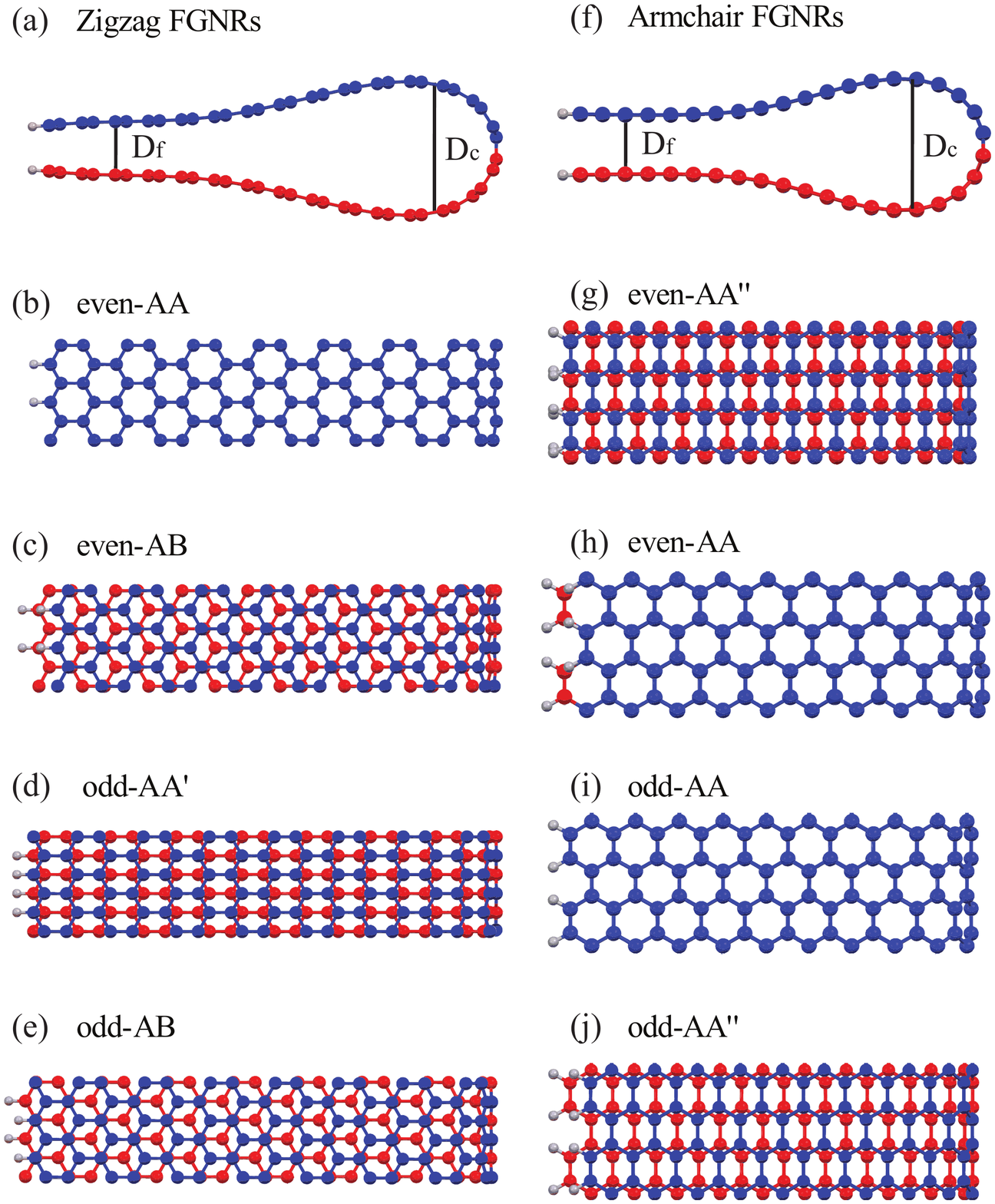}}
\caption{(a) Side views of even-$AA$ zigzag FGNRs. Top views of zigzag FGNRs with respect to (b) even-$AA$, (c) even-$AB$, (d) odd-$AA^{\prime}$, and (e) odd-$AB$. Side views of (f) odd-$AA$ armchair FGNRs. Top views of armchair FGNRs with respect to (g) odd-AA, (h) odd-$AA^{\prime\prime}$, (i) even-$AA$, and (j) odd-$AA^{\prime\prime}$.}
\label{Fig1}
\end{figure}

A FGNR, formed by folding a single flat ribbon, can be viewed as two flat ribbons connected by a fractional nanotube in the position to create the curved surface, as shown in Figs. 1(a) and 1(f). The open edges are passivated by hydrogen atoms (grey circles), and the width is characterized by the number of zigzag or dimer lines. Based on the ribbon width, edge structures, and stacking configuration in the flat region, there are eight types of stable FGNRs: (I) $AA$ and (II) $AB$ stackings with even zigzag lines, (III) $AA^{\prime}$ and (IV) $AB$ stackings with odd zigzag lines, (V) $AA^{\prime\prime}$ and (VI) $AA$ stackings with even dimer lines, and (VII) $AA^{\prime\prime}$ and (VIII) $AA$ stackings with odd dimer lines. In the $AB$ stacking, half of the atoms share identical (x,y) positions, while all carbons in the $AA$ stacking own the same projections. The $AA^{\prime}$ and $AA^{\prime\prime}$ stackings are, respectively, obtained by shifting the upper layer of the $AA$ stackings along the x-direction with displacements of $b/4$ and $\sqrt 3 b/2$ ($b$ is C-C bond length). Furthermore, carbon atoms on both layers have the same chemical environment. The hydrogen atoms in the Types (I), (II), (VII) and (VIII) are aligned in the same straight line; however, the others types are staggered on different lines.

\vskip 0.6 truecm
\par\noindent
{\bf 3. Optimal geometry and electronic properties}
\vskip 0.3 truecm

Each stacking configuration exhibits unique geometric properties. The curved and flat parts of a FGNR, separately, determine the nanotube diameter ($D_{c}$) and the interlayer distance ($D_{f}$) (Figs. 1(a) and 1(f)). The former represents the influence of the ribbon curvature, and the latter reflects the stacking effect. Both distances are significantly affected by the ribbon width and stacking configuration. As for zigzag FGNRs, the width-dependence of $D_{c}$ and $D_{f}$  are similar for the four types of stacking configurations, as shown in Fig. 2(a). $D_{c}$  initially grows with an increase of $N_{y}$ and then reaches a saturated value at $N_{y}\geq$ 40. For large $N_{y}$'s, $D_{c}$'s of the even-$AA$, even-$AB$, odd-$AA^{\prime}$ and odd-$AB$ stackings are, respectively, 9.2 ${\AA}$, 9.3 ${\AA}$, 8.7 ${\AA}$ and 8.5 ${\AA}$. Moreover, $D_{f}$'s are less affected by $N_{y}$. This weak dependence is reflected in the small variation of average $D_{f}$'s which are, respectively, 3.5 ${\AA}$, 3.2 ${\AA}$, 3.3 ${\AA}$ and 3.2 ${\AA}$ following the same sequence as for the $D_{c}$'s. For armchair systems (Fig. 2(b)), on the other hand, the $N_{y}$-dependence of $D_{c}$ is different from the above-mentioned relations. $D_{c}$ weakly depends on the ribbon width except for the even-$AA$ stacking, whose $D_{c}$ decreases with increasing $N_{y}$ and then reaches a final value at $N_{y}\geq$ 40. The $D_{c}$'s of the even-$AA$, even-$AA^{\prime\prime}$, odd-$AA$, and odd-$AA^{\prime\prime}$ stackings are, respectively, 7.1 ${\AA}$, 7.0 ${\AA}$, 7.0 ${\AA}$, and 7.2 ${\AA}$. The dependence of $D_{f}$ on the ribbon width is negligible, and the $D_{f}$ of the $AA$-stacked configuration is larger than that of the $AA^{\prime\prime}$-stacked one, regardless of the even or odd width. The perceivable $D_{f}$'s of the $AA$ and $AA^{\prime\prime}$ stackings are 3.5 ${\AA}$ and 3.3 ${\AA}$, respectively.

\begin{figure}[htbp]
\center
\rotatebox{0} {\includegraphics[width=14cm]{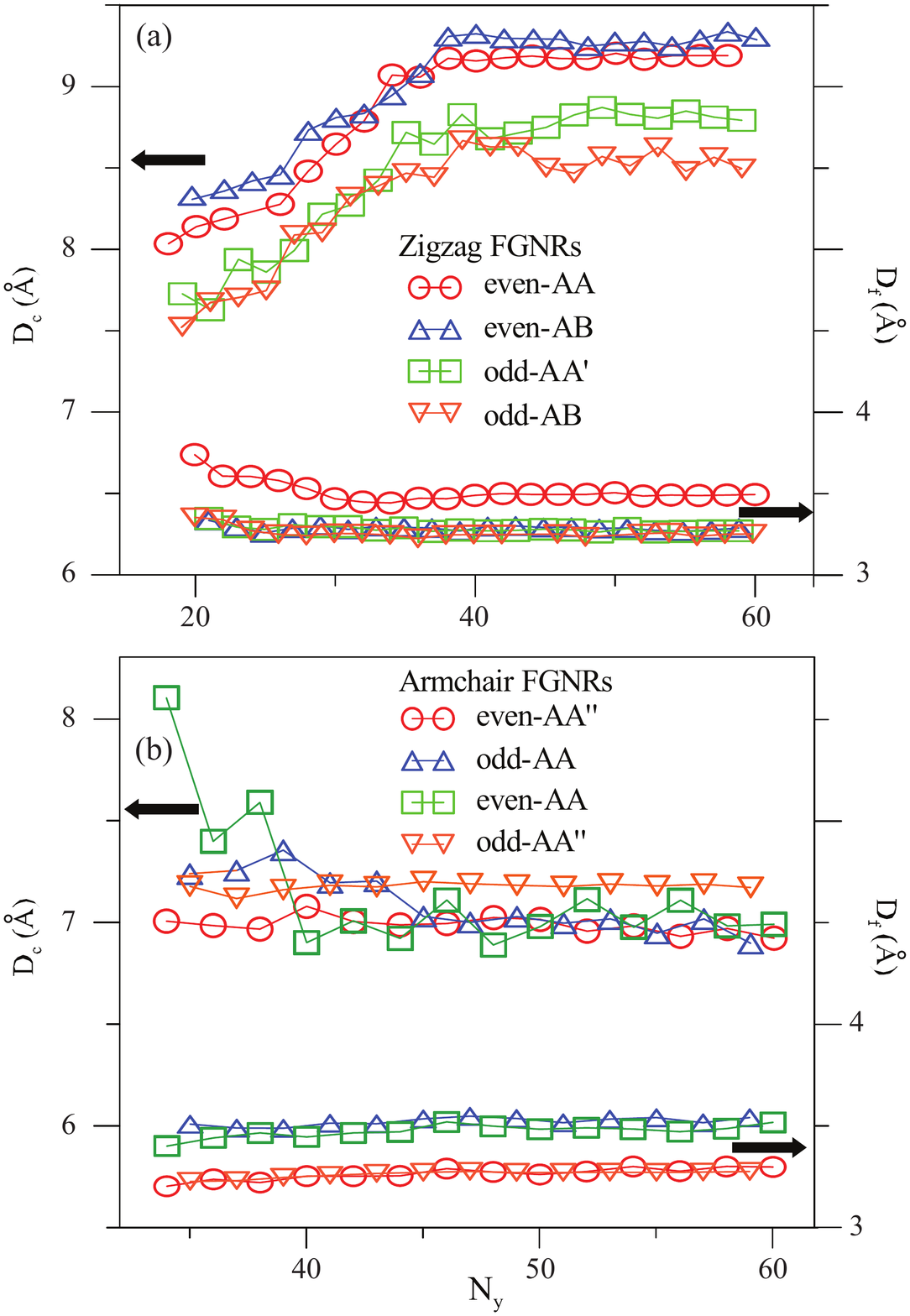}}
\caption{Width-dependence of $D_{c}$ and $D_{f}$ for different (a) zigzag FGNRs: even-$AA$ (red circles), even-$AB$ (blue triangles), odd-$AA^{\prime}$ (green squares), odd-$AB$ (orange triangles), and (b) armchair FGNRs: even-$AA^{\prime\prime}$ (red circles), odd-$AA$ (blue triangles), even-$AA$ (green squares), odd-$AA^{\prime\prime}$ (orange triangles).}
\label{Fig2}
\end{figure}

When the ribbon width is large enough, both $D_{f}$ and $D_{c}$ are nearly constant. $D_{f}$ is dominated by the stacking effect, meaning that even- and odd-$AB$ stackings have shorter interlayer distances. This result is in good agreement with those presented in bilayer graphene nanoribbons\cite{41} and bilayer graphenes\cite{28}. On the other hand, the fact that $D_{c}$ is shorter in armchair FGNRs than in zigzag ones implies that the former are subjected to a larger deformation. The deformation, which takes place tangent to the direction of the dimer or zigzag line, is responsible for this result. Similar deformations have also been observed in zigzag and armchair carbon nanotubes\cite{42}. Another important difference related to the curved structure is displayed in the formation energy.

The formation energy ($\Delta E$) is defined as the difference between the total energy of a FGNR and that of a flat GNR. This energy is dominated by the interlayer interaction, the edge-induced energy and the bending energy, which are mainly contributed by the flat two-layer interaction, the edge-edge interaction and the mechanical strain, respectively. The third energy is always larger than the sum of the first and second ones, so $\Delta E$ is greater than zero. Note that the edge-edge interactions are inversely proportional to the edge-edge distance ($d_{e}$). $\Delta E$ decreases with the increasing width, because the edge-edge interaction and the bending energy for a saturated $D_{c}$ are nearly unchanged, while the overlap area of two flat sheets grows in a widened ribbon. When the width is sufficiently wide ($N_y\geq$ 38 in Fig. 3(a))), $\Delta E$'s of zigzag systems exhibit the following descending order: even-$AA$ stacking $>$ odd-$AA^\prime$ stacking $>$ odd-$AB$ stacking $>$ even-$AB$ stacking. The $AB$-stacked FGNRs have a higher stability since they possess larger interlayer interactions. Moreover, the even-$AB$ stacking has a $\Delta E$ lower than the odd-$AB$ one. The $d_{e}$'s of both systems are, respectively, 3.1 ${\AA}$ and 3.6 ${\AA}$, which leads to the stronger edge-edge interaction in the even stacking systems. Such influence is also applied to narrow ribbons; therefore, $\Delta E$ associated with the even-$AA$ stackings ($d_{e}=$2.9${\AA}$) is lower than that of the odd-$AA^\prime$ ones ($d_{e}=$3.5 ${\AA}$). On the other hand, the edge-edge interactions do not play an important role in the $N_y$-dependence of $\Delta E$, but the relationship is dominated by the stacking configurations; $AA^{\prime\prime}$-stacked armchair FGNRs are more stable than $AA$-stacked ones, irrespective of the dimer line number being even or odd (Fig. 3(b)). The edge-edge interaction is indistinguishable for the four types of armchair FGNRs with almost the same $d_{e}$ ($\sim 3.6 {\AA}$). When the ribbon width is larger, the reason for the armchair FGNRs but not the zigzag ones having a higher $\Delta E$ can be explained by the fact that $D_{c}$ is shorter in the former than in the latter.

\begin{figure}[htbp]
\center
\rotatebox{0} {\includegraphics[width=14cm]{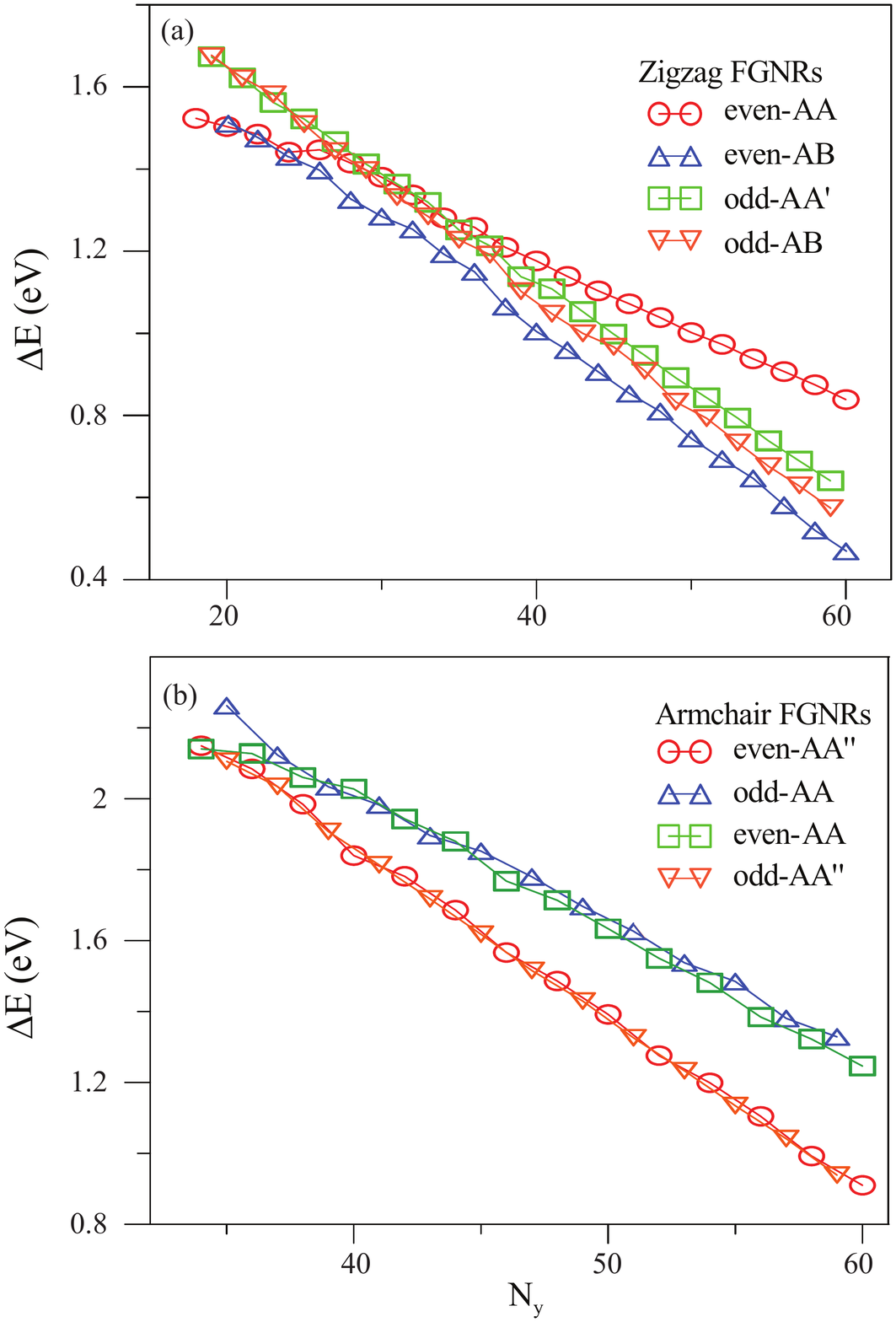}}
\caption{Variations of the width-dependent formation energies for (a) zigzag FGNRs, and (b) armchair FGNRs.}
\label{Fig3}
\end{figure}

For zigzag systems, the magnetic properties need to be taken into account. The spin arrangements can be changed by the geometric structure. The even-$AA$ and even-$AB$ configurations have no magnetic moments. However, the odd-$AA^{\prime}$ and odd-$AB$ systems, like monolayer GNR\cite{23}, present magnetic moments at the open edges. The main reason is that the edge-edge interactions can effectively suppress the magnetic moments at a sufficiently short distance ($<$3.5 ${\AA}$ )\cite{28}. In addition, the edge-edge distances for these four systems are, respectively, 2.9 ${\AA}$, 3.1 ${\AA}$, 3.5 ${\AA}$, and 3.6 ${\AA}$. The magnetic moments cannot clearly affect the above-mentioned properties, but their effects on electronic properties are strong. The systems without magnetic moments predictably have different electronic structures in comparison to monolayer GNR. Moreover, the electronic properties of the spin-configuration-dependent systems can be modulated by the geometric structure.

\begin{figure}[htbp]
\center
\rotatebox{0} {\includegraphics[width=14cm]{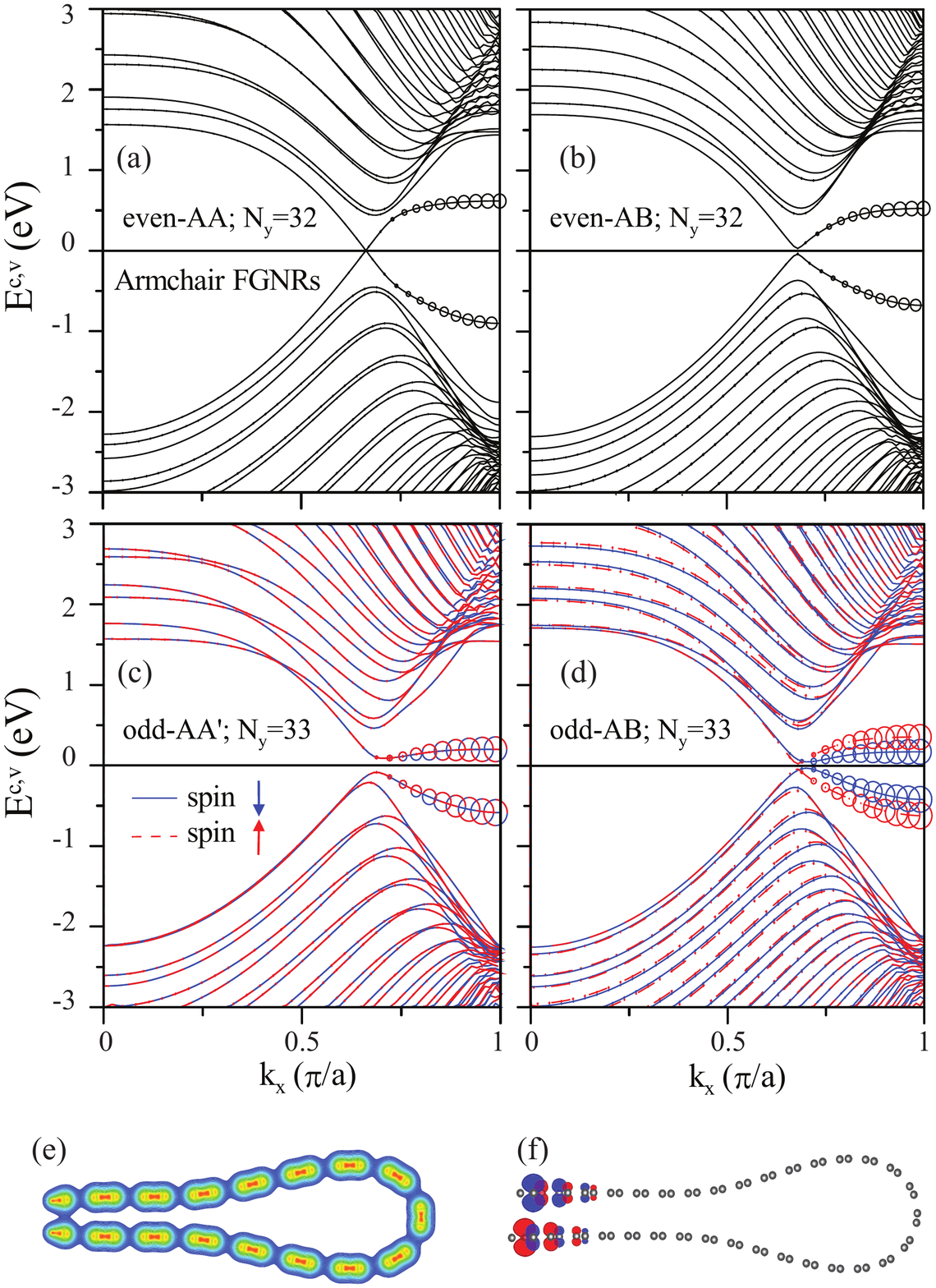}}
\caption{Band structures of (a) even-$AA$, (b) even-$AB$, (c) odd-$AA^{\prime}$; (d) odd-$AB$ stacked zigzag FGNRs. (e) Total charge density of even-$AA$ stacking. (f) Partial charge densities with spin arrangements of the odd-$AB$ stacking. The red color indicates spin up and the blue color spin down.}
\label{Fig4}
\end{figure}

The various 1D FGNRs present the feature-rich electronic properties. The band structures are significantly affected by the edge structure, stacking configuration, curvature effect and ribbon width. The zigzag FGNRs exhibit many 1D energy bands, as shown in Figs. 4(a)-4(d). The occupied ($E^{v}$) and unoccupied ($E^{c}$) energy bands are asymmetric to the Fermi level ($E_{F}$ = 0). Most of the subbands have parabolic energy dispersions. Each subband owns several band-edge states characterized by a local minimum or maximum. These states are located at $k_{x}=$ 0 and 1 (in units of periodical length), and some are located in between them. In the higher-energy region, there are many anti-crossing subbands with band-edge states close to $k_{x}=$1, mainly owing to the interlayer interactions and the enhanced hybridization of the four orbitals ($2s, 2p_{x}, 2p_{y}, 2p_{z}$) on the curved surface. In the low-lying subbands the pair of subbands nearest to $E_{F}$ in the regions of $2/3\leq k_{x}\leq1$ and $0\leq k_{x}\leq 2/3$, respectively, possess partially flat and parabolic dispersions. The former are mainly contributed by the local edge atoms (the circle radii representing the contributions of the edge atoms). These two subbands are quite different among the four types of zigzag FGNRs. The even-$AA$ stacked systems belong to a 1D metal, while the other three systems are direct-gap semiconductors. The valence and conduction bands in the first systems intersect each other at $E_{F}=0$, and the Fermi-momentum state is similar to the Dirac point in an armchair carbon nanotube\cite{43}. This is because the sufficiently strong edge-edge interactions induces more charge carriers which appear in the interlayer region near the boundary and thus lead to the loop-like charge density (Fig. 4(e)). For the even-$AB$ stackings, the linearly intersecting bands become two anti-crossing parabolic bands with a narrow direct energy gap, as shown in Fig. 4(b). The metal-semiconductor transition is caused by the variation of the stacking configuration, which is responsible for the fact that a distinct geometric symmetry can open the linearly intersecting bands in the same way as in a bilayer graphene nanoribbon\cite{28} and a large  closed carbon nanotube\cite{40}. On the other hand, the odd-$AA^{\prime}$ and odd-$AB$ FGNRs, with magnetism, exhibit two spin-dependent band structures (Figs. 4(c) and 4(d)). Both spin-up and spin-down subbands are doubly degenerate for the former, while the spin degeneracy is absent in the latter. The odd-$AB$ systems do not have the mirror symmetry about the z-axis so that the magnetic environments are different for edge atoms with spin-up and spin-down configurations (Fig. 4(f)). However, the odd-$AA^{\prime}$ stackings exhibit the opposite behavior. This can account for the spin splitting or degeneracy of the energy subbands. Among the semiconducting zigzag FGNRs, the odd-$AA^{\prime}$ ones have the largest energy gaps. In addition, the energy spacing at $k_{x}=$1 is smaller in the odd-systems than in the even ones.

\begin{figure}[htbp]
\center
\rotatebox{0} {\includegraphics[width=14cm]{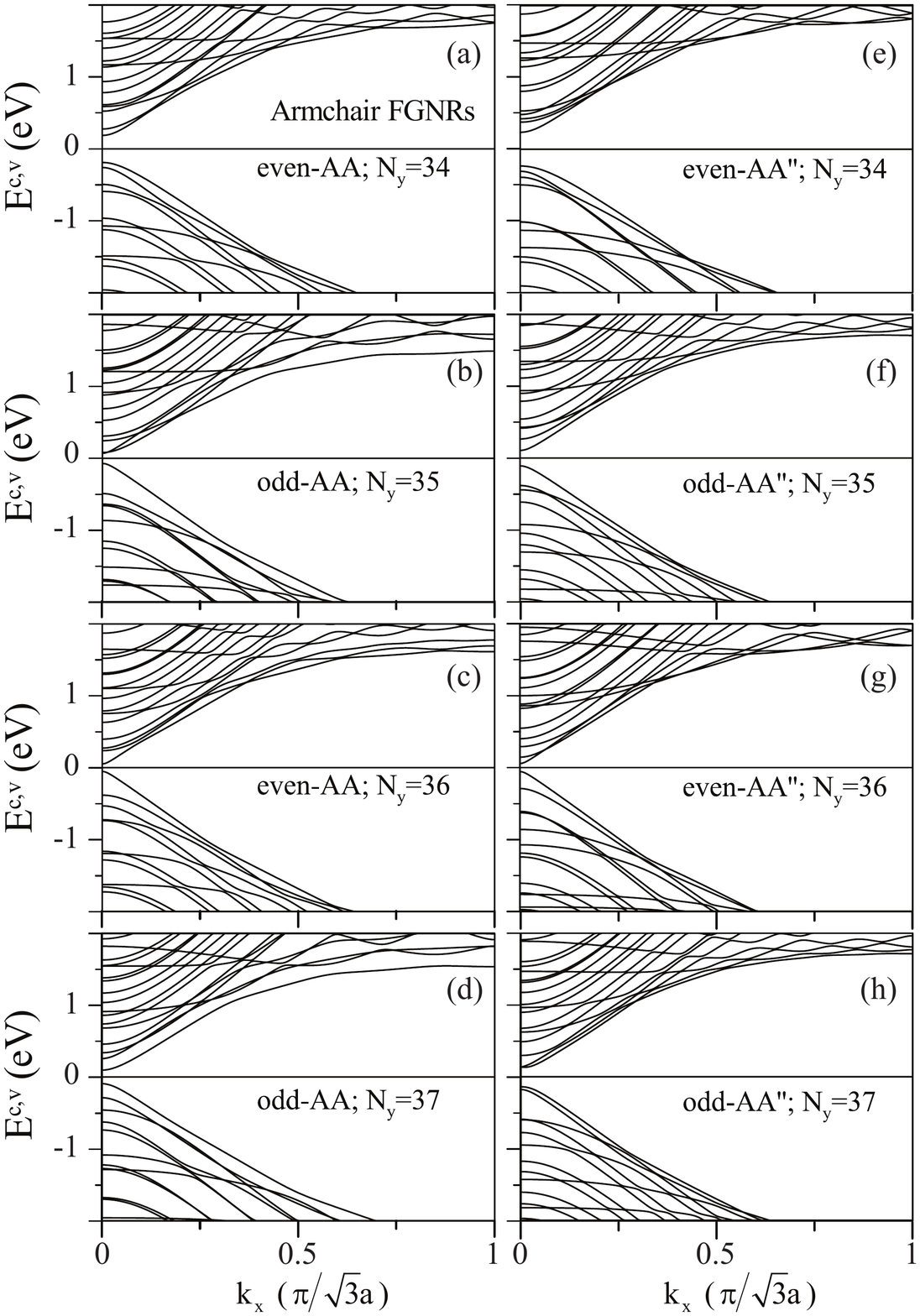}}
\caption{Width-dependent band structures of $AA$ stacked armchair FGNRs: (a)$N_{y}=34$, (b)$N_{y}=35$ , (c) $N_{y}=36$; (d) $N_{y}=37$. The ribbon-dependent band structures of $AA^{\prime\prime}$ stacked armchair FGNRs: (e)$N_{y}=34$, (f)$N_{y}=35$, (g)$N_{y}=36$; (h)$N_{y}=37$.}
\label{Fig5}
\end{figure}

Compared to the zigzag FGNRs, the armchair ones display an entirely different type of band structure. Energy dispersions of the armchair FGNRs are almost independent of the stacking configuration, as shown in Figs. 5(a)-5(h). In the region of $|E^{c,v}|\leq1$ eV, each subband exhibits one band-edge state at $k_{x}=0$. However, some of the band-edge states are doubly degenerate due to the stacking effect. All the armchair FGNRs belong to direct-gap semiconductors, being determined by the $k_{x}=0$ states. The energy gaps are sensitive to changes in the ribbon width and the stacking configuration. The $N_{y}=34$ even-$AA$ and $N_{y}=37$ odd-$AA$ stackings possess larger energy gaps. Similar results have be predicted for the flat nanoribbon case, where the largest energy gap is from the group of $N_{y}=3p+1$; that is attributed to the finite-width quantum confinement\cite{23}. When the stacking configuration changes, the $N_{y}=34$ even-$AA^{\prime\prime}$ and $N_{y}=37$ odd-$AA^{\prime\prime}$ stacking FGNRs also own larger energy gaps. In addition, the $AA^{\prime\prime}$ stacking has larger gaps than those possessed by the $AA$ stacking.

\begin{figure}[htbp]
\center
\rotatebox{0} {\includegraphics[width=14cm]{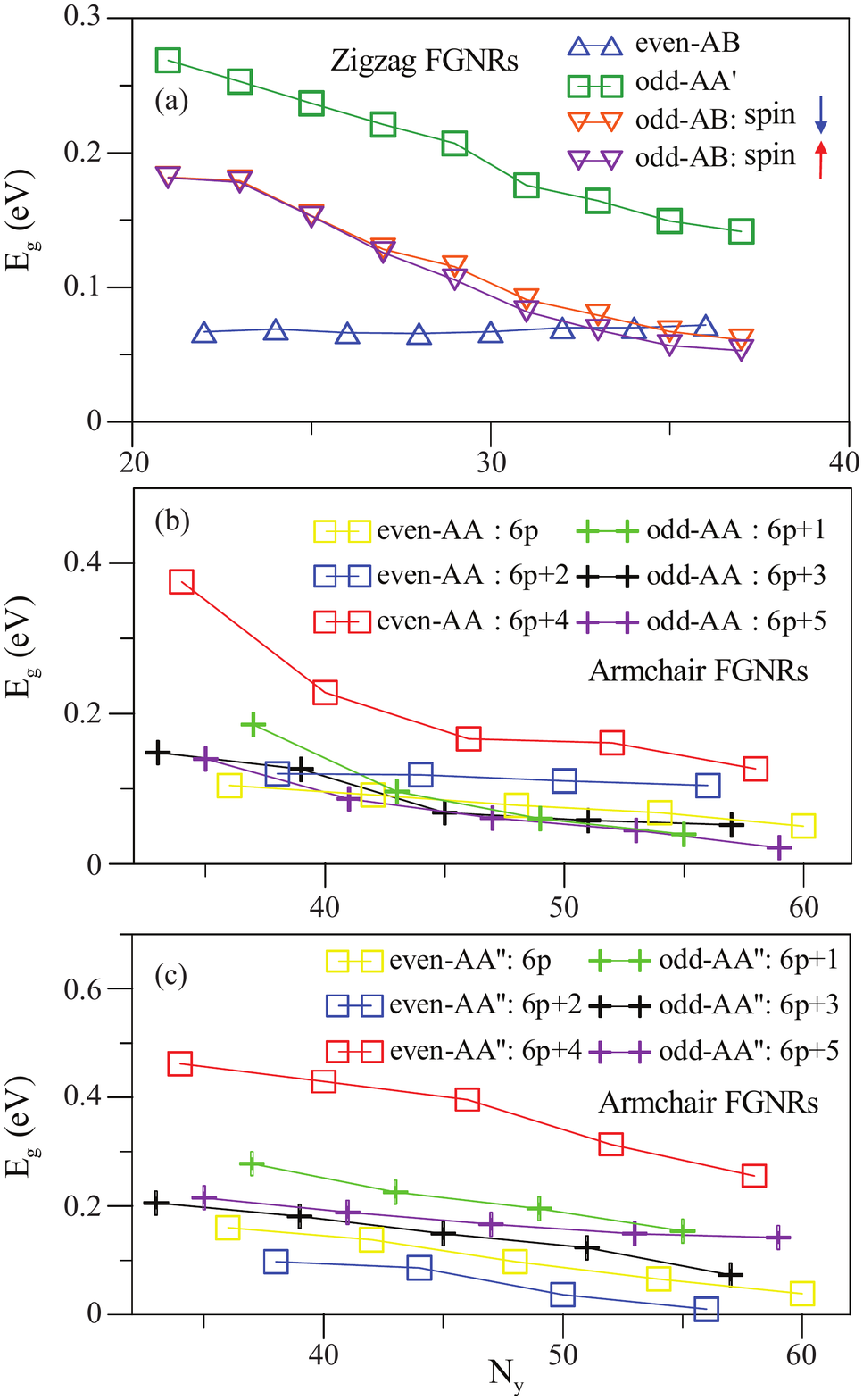}}
\caption{Variation of width-dependent energy gaps of (a) zigzag, (b) $AA$ stacked armchair and (c) $AA^{\prime\prime}$ stacked armchair FGNRs.}
\label{Fig6}
\end{figure}

The versatile width dependence of the energy gaps in various FGNRs deserve closer examination. For the zigzag systems, $E_{g}$ decreases with an increasing ribbon width. The energy gaps are strongly dependent on the stacking and curvature effects; therefore, small zigzag FGNRs exhibit the following descending order: odd-$AA^{\prime}$ stacking $>$ odd-$AB$ stacking with spin down $>$ odd-$AB$ stacking with spin up $>$ even-$AB$ stacking. The dependence of $E_{g}$ on $N_{y}$ is weak for the even-$AB$ stackings, while the opposite is true for the odd-$AA^{\prime}$ and odd-$AB$ ones (Fig. 6(a)). This is because the earlier explanation on the formation of energy gap in the even-$AB$ stackings considerably differs from what accounts for the odd-$AA^{\prime}$ and odd-$AB$ ones. On the other hand, the energy gaps of armchair FGNRs can be further divided into six groups, based on the two types of stacking configurations. $E_{g}$'s corresponding to the $AA$ and $AA^{\prime\prime}$ stackings, respectively, possess three $N_{y}$-relationships, all of which decline with a larger ribbon width as shown in Fig. 6(b). The even-$AA$ stacking presents the largest (smallest) energy gap, when the ribbon width is $N_{y}=6p+4$ ($6p$). Moreover, the odd-$AA$ stacking owns a smaller gap in $N_{y}=6p+5$, but the largest energy gap lies in $N_{y}=6p+3$ or $6p+1$. Similarly, the even-$AA^{\prime\prime}$ and odd-$AA^{\prime\prime}$ stackings are each classified into one of three width-dependent energy gap groups. The even-$AA^{\prime\prime}$ stackings belong to the largest energy gap among the discussed armchair system when the ribbon width corresponds to the $6p+4$ group. The smallest energy gap of the even-$AA^{\prime\prime}$ stackings is observed in the $6p+2$ group. The odd-$AA^{\prime\prime}$ stacking for the $6p+1$ ($6p+3$) group owns a larger (smaller) energy gap. These complex width-dependences are different from those presented in flat nanoribbons, mainly owing to the combined stacking and curvature effects in this unique folding structure. The ribbon width corresponding to the smallest energy gap is the $3p+2$ group for the flat nanoribbons; however, this is not exactly the case for the FGNRs. Instead, another group associated with the smallest energy gap here is $3p$, for example, the even-$AA$ and odd-$AA^{\prime\prime}$ stackings.

The DOS directly reflects the main characteristics of the band structures. The low-energy DOS of the even-$AA$ stacked zigzag FGNRs, as shown in Fig. 7(a) by the black solid curve, exhibits a plateau near $E_{F}$, two prominent peaks (solid and hollow squares), and many square-root asymmetric peaks, respectively, coming from the intersecting linear bands, the partially rounded bands, and the parabolic bands (Fig. 4(a)). The constant plateau indicates the metallic behavior, as seen in a 1D armchair carbon nanotube\cite{43}. The two prominent peaks on either side of $E_{F}$ are mainly contributed by the local edge atoms. However, in the even-$AB$ stacking, the continuous plateau of the DOS is destroyed and evolves into two separate peaks (red solid and hollow circles in Fig. 7(a)), which determine a direct narrow gap. The odd-$AA^{\prime}$ and odd-$AB$ stackings possess two DOS's: one is associated with spin up (red curve) and the other is related to spin down (blue curve). The contributions are not distinguishable to DOS from the spin-up and spin-down states for the odd-$AA^{\prime}$ stackings (Fig. 7(b)), but are distinct for the odd-$AB$ ones (Fig. 7(c)). Moreover, regardless of the spin states, the energy spacings of the two adjacent prominent peaks (solid and hollow squares) are smaller in the odd-$AA^{\prime}$ and odd-$AB$ systems than in the even-$AB$ ones. The energy gap of the two separate peaks nearest to $E_{F}$ is the largest (smallest) in the odd-$AA^{\prime}$ stackings (even-$AB$ stackings). The low-lying features in DOS are dramatically changed by the edge structure, as displayed for the armchair FGNRs in Figs. 7(d)-7(g). The peak energies, numbers and heights near $E_{F}$ contrast sharply with those in the zigzag systems. There is no pair of prominent peaks coming from the edge atoms; on the contrary, many peaks originate from the $k_{x}=0$ band-edge states. There are some relatively high peaks (green diamonds) are exhibited, mainly owing to the double degeneracy.

\begin{figure}[htbp]
\center
\rotatebox{0} {\includegraphics[width=14cm]{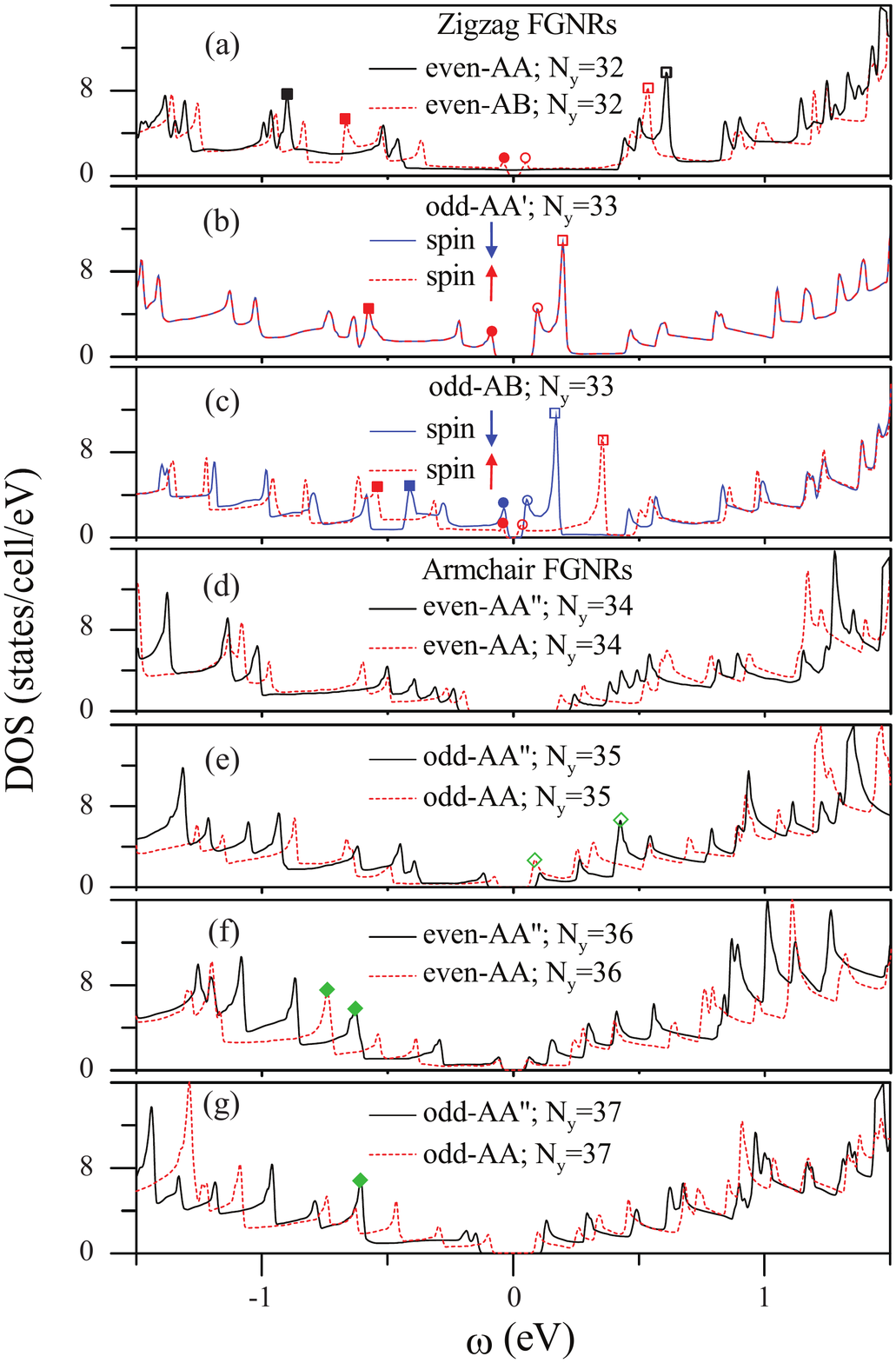}}
\caption{Density of states for (a) even-$AA$ and even-$AB$, (b) odd-$AA^{\prime}$, (c) odd-$AB$ stacked zigzag FBNRs, and for $AA$ and $AA^{\prime\prime}$ stacked armchair FGNRs with (d) $N_{y}=34$, (e) $N_{y}=35$, (f) $N_{y}=36$; (g) $N_{y}=37$.}
\label{Fig7}
\end{figure}

The curvature, edge, and stacking effects, which play an important role in the energy, number, and height of peaks in DOS, can be verified by the scanning tunneling spectroscopy measurements. The tunneling conductance is roughly proportional to the DOS as a measurement reflecting the major structures. The various features of the DOS can be used to identify the type of FGNR. The zigzag FGNRs possess a pair of prominent peaks on either side of $E_{F}$, a characteristic not reflected in the armchair systems. The DOS associated with the electron spin is the key to differentiate whether the width of the zigzag systems is even or odd. Only the even-$AA$ stacking has free-carrier states at $E_{F}$. The splitting of the spin degeneracy induces more peaks for the odd-$AB$ stackings. For the armchair FGNRs, the odd-$AA$, even-$AA^{\prime\prime}$, even-$AA$, and odd-$AA^{\prime\prime}$ stackings cannot be easily identified due to the similar DOS. However, the energy gaps in DOS are a critical property in confirming the width-dependent six groups. In previous experimental measurements, the curvature effect in carbon nanotubes\cite{44} and rippled graphenes\cite{45} and the stacking effect in few-layer graphenes\cite{46} have been verified. As expected, these two combined effects in FGNRs can be further examined by the STS measurements.

\vskip 0.6 truecm
\par\noindent
{\bf 4. Conclusion}
\vskip 0.3 truecm

The geometric and electronic properties of zigzag and armchair FGNRs are thoroughly investigated by calculations based on ab initio functional theory. Eight types of stacking configurations and various ribbon widths are considered to reveal the fundamental properties, including optimal geometries, formation energies, charge densities, band structures, energy gaps and DOS. FGNRs show many important features, such as the constant $D_{c}$ and $D_{f}$ for larger ribbon widths, the destruction or creation of magnetism, the lower formation energies in the $AB$ or $AA^{\prime\prime}$ stackings, the metal-semiconductor transitions, and the monotonous width-dependence of the energy gap. The predicted results could be verified by experimental measurements of scanning tunneling currents\cite{47}, optical spectra\cite{48}, and transport properties\cite{49}.

The geometric structure of FGNR can be divided into the flat and curved parts; therefore, the fundamental properties are strongly dependent on the competition or cooperation among the stacking, curvature and edge effects. The zigzag even-$AB$ stackings or armchair $AA^{\prime\prime}$ stackings are the most stable stacking configurations. The electronic structures of FGNR in the higher-energy region possess anti-crossing bands because of the interlayer interactions and hybridization of four orbitals on the curved surface, while in the lower-energy region the energy gap and energy spacing of the band-edge states are sensitive to ribbon width and stacking configuration. The zigzag even-$AA$ stacking possesses intersecting linear bands and the zigzag even-$AB$ one is a narrow-gap semiconductor. The odd-$AA^{\prime}$ and odd-$AB$ stackings are magnetic materials. The latter has two energy gaps associated with the spin-up and spin-down states. All armchair FGNRs are direct-gap semiconductors. These energy gaps exhibit a monotonic relationship, which decreases with increased ribbon width. It is worth noting that the armchair systems can be classified into six groups of width-dependent energy gaps. The width-dependencies in the armchair FGNRs are different from those in flat nanoribbons. In other words, the smallest energy gap of the FGNRs is not necessarily associated with the $N_{y}=3p+2$ group. The DOS features, including the form, number, intensity and energy of the special structures, directly reflect the unique electronic properties. These rich electronic properties and various energy gaps are promising for the potential application in electronic devices.

\par\noindent {\bf Acknowledgments}

We thank the Physics Division, National Center for Theoretical Sciences (South), and the Nation Science Council of Taiwan under Grants NSC 102-2112-M-182-002-MY3 and NSC 102-2112-M-006-007-MY3 for monetary support.

\newpage
\renewcommand{\baselinestretch}{0.2}

\newpage \centerline {\Large \textbf {FIGURE CAPTIONS}}

\vskip0.5 truecm 

Fig. 1 - (a) Side views of even-$AA$ zigzag FGNRs. Top views of zigzag FGNRs with respect to (b) even-$AA$, (c) even-$AB$, (d) odd-$AA^{\prime}$, and (e) odd-$AB$. Side views of (f) odd-$AA$ armchair FGNRs. Top views of armchair FGNRs with respect to (g) odd-AA, (h) odd-$AA^{\prime\prime}$, (i) even-$AA$, and (j) odd-$AA^{\prime\prime}$.

Fig. 2- Width-dependence of $D_{c}$ and $D_{f}$ for different (a) zigzag FGNRs: even-$AA$ (red circles), even-$AB$ (blue triangles), odd-$AA^{\prime}$ (green squares), odd-$AB$ (orange triangles), and (b) armchair FGNRs: even-$AA^{\prime\prime}$ (red circles), odd-$AA$ (blue triangles), even-$AA$ (green squares), odd-$AA^{\prime\prime}$ (orange triangles).

Fig. 3 - Variations of the width-dependent formation energies for (a) zigzag FGNRs, and (b) armchair FGNRs.

Fig. 4 - Band structures of (a) even-$AA$, (b) even-$AB$, (c) odd-$AA^{\prime}$; (d) odd-$AB$ stacked zigzag FGNRs. (e) Total charge density of even-$AA$ stacking. (f) Partial charge densities with spin arrangements of the odd-$AB$ stacking. The red color indicates spin up and the blue color spin down.

Fig. 5 - Width-dependent band structures of $AA$ stacked armchair FGNRs: (a)$N_{y}=34$, (b)$N_{y}=35$ , (c) $N_{y}=36$; (d) $N_{y}=37$. The ribbon-dependent band structures of $AA^{\prime\prime}$ stacked armchair FGNRs: (e)$N_{y}=34$, (f)$N_{y}=35$, (g)$N_{y}=36$; (h)$N_{y}=37$.

Fig. 6 - Variation of width-dependent energy gaps of (a) zigzag, (b) $AA$ stacked armchair and (c) $AA^{\prime\prime}$ stacked armchair FGNRs.

Fig. 7 - Density of states for (a) even-$AA$ and even-$AB$, (b) odd-$AA^{\prime}$, (c) odd-$AB$ stacked zigzag FBNRs, and for $AA$ and $AA^{\prime\prime}$ stacked armchair FGNRs with (d) $N_{y}=34$, (e) $N_{y}=35$, (f) $N_{y}=36$; (g) $N_{y}=37$.

\end{document}